\documentclass[conference]{IEEEtran}
\IEEEoverridecommandlockouts
\usepackage{cite}
\usepackage{amsmath,amssymb,amsfonts}
\usepackage{algorithmic}
\usepackage{graphicx}
\usepackage{textcomp}
\usepackage{xcolor}
\def\BibTeX{{\rm B\kern-.05em{\sc i\kern-.025em b}\kern-.08em
    T\kern-.1667em\lower.7ex\hbox{E}\kern-.125emX}}

\definecolor{bluegray}{rgb}{0.4, 0.6, 0.8}

\definecolor{darkgreen}{RGB}{0,120,0}
\definecolor{darkblue}{RGB}{0,0,200}
\definecolor{airforceblue}{rgb}{0.36, 0.54, 0.66}
\definecolor{bleudefrance}{rgb}{0.19, 0.55, 0.91}
\definecolor{cerulean}{rgb}{0.0, 0.48, 0.65}
\newcommand{\green}[1]{{\color{black}#1}}

\begin{document}

\title{Beyond the Bridge: Contention-Based Covert and Side Channel Attacks on Multi-GPU Interconnect}

\author{{Yicheng Zhang$^1$,
Ravan Nazaraliyev$^1$,
Sankha Baran Dutta$^2$,
Nael Abu-Ghazaleh$^1$,
Andres Marquez$^2$,
Kevin Barker$^2$
}\\

$^1$University of California, Riverside\\
$^2$Pacific Northwest National Laboratory
}


\maketitle

\begin{abstract}
High-speed interconnects, such as NVLink, are integral to modern multi-GPU systems, acting as a vital link between CPUs and GPUs. This study highlights the vulnerability of multi-GPU systems to covert and side channel attacks due to congestion on interconnects. An adversary can infer private information about a victim's activities by monitoring NVLink congestion without needing special permissions. Leveraging this insight, we develop a covert channel attack across two GPUs with a bandwidth of 45.5 kbps and a low error rate, and introduce a side channel attack enabling attackers to fingerprint applications through the shared NVLink interconnect. 
\end{abstract}


\section{Introduction}
Graphics Processing Units (GPUs) are crucial for accelerating various applications, including computer vision~\cite{sc_workshop_zhang}, Extended Reality (XR)~\cite{zhang2023s}, language models~\cite{li2023rt}, health care~\cite{lai2024language}, and more. As dataset sizes grow, experiments on a single GPU may take days, prompting the use of multi-GPU systems to speed up these applications. Security and privacy concerns within the realm of both CPU and GPU have garnered increasing attention. Conventional I/O interconnects such as PCIe have been demonstrated to be susceptible to side channel attacks~\cite{tan2021invisible, dutta2021leaky, side2022lockeddown}. This study highlights that contemporary high-speed multi-GPU interconnects are also vulnerable to covert and side channel attacks stemming from congestion on these links. By monitoring the data transfer rates of the shared NVLink protocol, adversaries can glean information about other users' activities. Specifically, we devise a covert channel attack across two GPUs, achieving a notable bandwidth. Additionally, we introduce a side-channel attack that allows attackers to profile applications running on multiple GPUs.

In this wild and emerging ideas (WEI) paper, we present the initial outcomes of our attacks on Nvidia multi-GPU systems. This paper introduces the \textbf{first} attack on multi-GPU interconnects using congestion timing leakages. 

\section{Background and Threat Model}
\label{sec:Background}

NVLink, developed by Nvidia~\cite{nvlink_whitepaper}, is a high-speed interconnect designed to enable rapid data exchange between CPUs and GPUs. It supports efficient read and write operations on the host and device memory. Each bidirectional link consists of two sublinks, ensuring high-bandwidth communication. Our experiments focus on the second version of NVLink,  supported by Tesla V100 GPUs~\cite{choquette2018volta}.

\noindent \textbf{{Threat model.}} Previous microarchitectural attacks~\cite{naghibijouybari2018rendered,side2022lockeddown,wei2020leaky} necessitated the co-location of victim and spy users on a single GPU. In contrast, our approach eliminates this requirement. Spy and victim users can share the same NVLink without being located on the same GPU or needing specialized system support. Additionally, the attacker does not require any special privileges or specialized system support.

\section{Cross GPU Covert Channel Attack}
\begin{figure*}[ht]
    \centering
    \includegraphics[width=\textwidth]{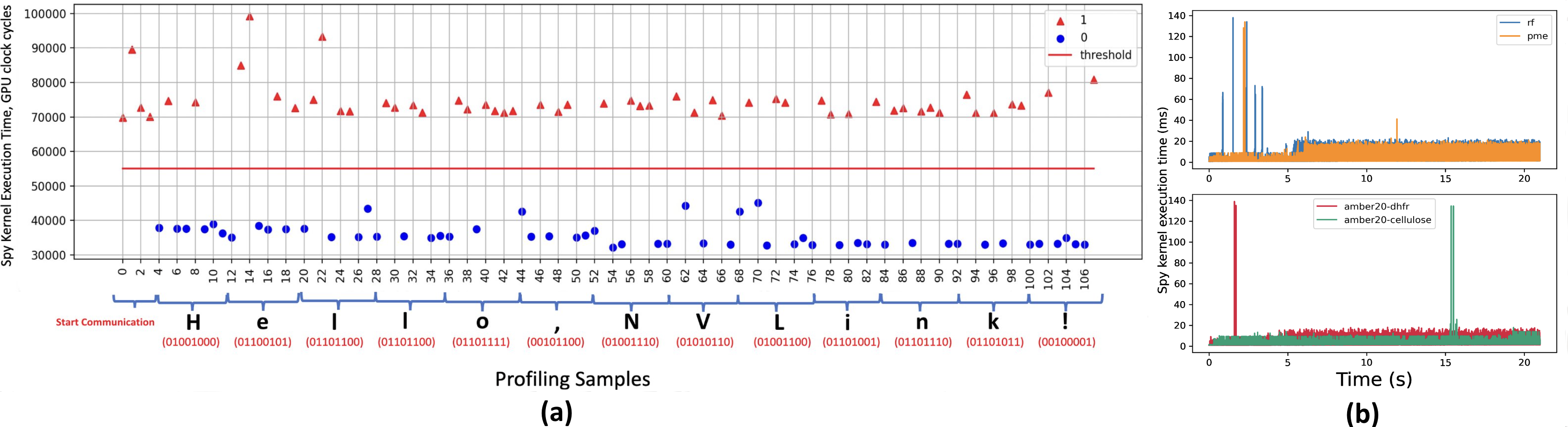}
    \caption{Two attacks in this work: (a) Cross-GPU covert message; (b) Application fingerprinting via NVLink congestion side channel leakage.
    }
    \label{fig:two_attacks}
\end{figure*}

We set up a scenario involving two Tesla V100 GPUs connected via NVLink. \green{The sender and receiver programs are situated on GPU0 and GPU1, respectively. The sender begins by allocating two 1.25 MB memory buffers: one on the remote GPU1 and another on the local GPU0. To transmit a '1', the sender executes a {\em cudaMemcpyPeer()} operation, transferring data from GPU1 to GPU0; to indicate a '0', it remains idle. Likewise, the receiver sets up two 256-byte memory buffers on the remote GPU0 and the local GPU1. It then initiates a {\em cudaMemcpyPeer()} operation to prompt data transfer from GPU0 to GPU1, simultaneously tracking the operation's execution time. The congestion within the GPU interconnects leads to differentiable profiling times: 28,356 clock cycles for '0' and 68,368 for '1'. A threshold of 55,000 clock cycles is employed to distinguish between bit '0' and '1'.

Fig.\ref{fig:two_attacks} (a) depicts the covert message transmission process. Communication begins upon the receiver's detection of four consecutive "1" bits, at which point the text-to-binary encoded message "Hello,NVLink!" is covertly transmitted. We evaluated the communication bandwidth and error rate by executing over 5 runs of 10000-bit message transmission. The measured bandwidth was recorded as 45.5 kbps, accompanied by an error rate of 3.22\%.}

\section{Cross GPU Side Channel Attack}

In this attack, a background spy application continuously monitors NVLink data transfer latency caused by other users' activities. The NVLink traffic traces are then analyzed to correlate with the victim's actions, allowing us to infer the applications the victim runs.

We demonstrate that an attacker can identify the specific HPC application being executed by eavesdropping on the NVLink timing side channel leakages. To illustrate this, we utilized four applications from the OpenMM~\cite{eastman2017openmm} benchmark as the victim applications, as listed below. We conducted experiments on two Nvidia Telsa V100s with an NVLink connection. The background spy program performs the following operations: it first allocates two chunks of memory on a remote GPU1 and local GPU0, then executes the vector addition kernel to force sending data from GPU1 to GPU0. Afterward, we use the timer function to record the execution time of the spy kernel and use it to detect the congestion influenced by victim programs.

\noindent \textbf{{OpenMM benchmark.}} OpenMM is a high-performance toolkit tailored for molecular dynamics simulations, supporting multi-GPU systems. We focus on four benchmark applications: rf, pme, amber20-dhfr, and amber20-cellulose.

\noindent \textbf{{Observing benchmark distinguishability.}} Fig.~\ref{fig:two_attacks} (b) showcases traces for four benchmarks: rf (blue), pme (orange), amber20-dhfr (red), and amber20-cellulose (green), highlighting their distinguishable characteristics. For future endeavors, we intend to employ standard machine learning classifiers, as done in prior studies~\cite{dutta2023spy,zhang2021stealing}, to further distinguish between these applications.

\section{Conclusion}
This paper explores congestion-based covert and side channel attacks on multi-GPU systems. We advocate for heightened awareness within our community regarding implementing multi-GPU interconnects with enhanced security measures.

\section*{Acknowledgment}
We thank our anonymous reviewers for their valuable comments. This work was supported by the U.S. DOE Office of Science, Office of Advanced Scientific Computing Research, under awards 66150: "CENATE - Center for Advanced Architecture Evaluation" and 76125: “AMAIS - Advanced Memory to support Artificial Intelligence for Science.” The Pacific Northwest National Laboratory is operated by Battelle for the U.S. Department of Energy under contract DE-AC05-76RL01830. The work was also partially supported by US National Science Foundation grants CNS-1955650 and CNS-2053383.

{
\bibliographystyle{IEEEtran}
{ \bibliography{main}}}

\end{document}